Spin-dependent tunneling through high-k LaAlO$_3$

V. Garcia

Institut des NanoSciences de Paris, Universités Paris 6 et Paris 7, CNRS UMR 7588, 140 rue de Lourmel 75015 Paris, France and Unité Mixte de Physique CNRS-Thales, Domaine de Corbeville, 91404 Orsay, France

M. Bibes*

Institut d'Electronique Fondamentale, Université Paris-Sud, 91405 Orsay, France

J.-L. Maurice, E. Jacquet, K. Bouzehouane, J.-P. Contour and A. Barthélémy

Unité Mixte de Physique CNRS-Thales, Domaine de Corbeville, 91404 Orsay, France

Abstract

We report on the use of the LaAlO$_3$ (LAO) high-k dielectric as a tunnel barrier in magnetic tunnel junctions. From tunnel magnetoresistance (TMR) measurements on epitaxial La$_{2/3}$Sr$_{1/3}$MnO$_3$/LAO/La$_{2/3}$Sr$_{1/3}$MnO$_3$ junctions, we estimate a spin polarization of 77% at low temperature for the La$_{2/3}$Sr$_{1/3}$MnO$_3$/LAO interface. Remarkably, the TMR of La$_{2/3}$Sr$_{1/3}$MnO$_3$/LAO/Co junctions at low bias is negative, evidencing a negative spin-polarization of Co at the interface with LAO, and its bias dependence is very similar to that of La$_{2/3}$Sr$_{1/3}$MnO$_3$/STO/Co junctions. We discuss possible reasons for this behaviour.

An important issue for the future of electronics is to combine the spin degree of freedom with conventional semiconductor functionalities (like carrier density control via doping or gate voltage) [1]. An obvious pre-requisite is the successful injection of a spin-polarized current into a semiconductor, which is not a trivial task. Indeed, in the simplest approach based on spin-injection from a ferromagnetic metal, the resistance mismatch between the semiconductor and the magnetic electrode impedes the efficient injection of spin-polarized carriers in the semiconductor. In principle, this problem can be circumvented by inserting a spin-conserving resistance at the metal-insulator interface [2]. This is for instance a thin insulating layer acting as a tunnel barrier [3]. At this point, several material-related difficulties may arise, as the structural and electronic properties of the barrier and of the barrier-semiconductor interface must be very well controlled.

On this particular problem, one can expect to benefit from the experience gained in the field of high-k dielectrics [4]. These highly insulating compounds are currently being considered to substitute $SiO_2$ as the gate material in future MOSFET structures. In the last few years, a huge experimental effort has been made to grow optimised layers of high-k dielectrics on Si. Most of them are binary oxides, such as $ZrO_2$ or $HfO_2$, but one of the more promising candidates is the $LaAlO_3$ perovskite (LAO) [5].

LAO has a high dielectric constant (~25 [6]), a large band-gap (5.1 eV [7]) and has been grown as amorphous [8] and epitaxial layers [9] on Si. Compared to other possible high-k perovskites that have been epitaxially grown on Si, like $SrTiO_3$ (STO) [10], LAO has an important advantage: its band-offset with Si is 1.8 eV [11], while that of STO is virtually zero [12], which disqualifies STO as a possible gate material and questions its potential for spin-injection. Being a perovskite ($ABO_3$), LAO is also structurally compatible with many functional compounds, like manganites, superconductors and ferroelectrics.

To use the same material as gate dielectric and tunnel barrier would certainly be an advantage for the fabrication and processing of future spintronics architectures. But before that, one has to know better the potential of LAO as a tunnel barrier for efficient spin-injection. This can be done by studying magnetic tunnel junctions based on LAO barriers, which is the purpose of this Letter. We describe the growth of tunnel junctions using 2.8 nm epitaxial films of LAO as barriers, and Co or $La_{2/3}Sr_{1/3}MnO_3$ (LSMO) as magnetic electrodes. Spin-dependent transport measurements yield a tunnel magnetoresistance (TMR) of 300% at low-temperature in LSMO/LAO/LSMO junctions, and of -19% for LSMO/LAO/Co junctions. This negative TMR indicates a negative spin-polarization of Co at the interface with LAO, as occurs with STO. The possible reasons for this behavior are discussed.

LSMO/LAO and LSMO/LAO/LSMO heterostructures have been grown by pulsed laser deposition using a stoichiometric ceramic target for LSMO and a single-crystalline target for LAO. (001)-oriented STO single crystals were used as substrates. The deposition temperature was set to 720°C and the pressure to 350 mTorr for LSMO and LAO. The laser rate was set to 2.5 Hz. At the end of deposition, the oxygen pressure was set to 300 Torr and kept constant during cooling. The LSMO/LAO/LSMO structure was capped in-situ by a Au layer 30 nm thick. The LSMO/LAO bilayer was transferred to a sputtering chamber. Before deposition of the 50 nm Co layer, a short oxygen plasma was applied to clean the LAO surface. This is a standard procedure that was already employed, for instance, to fabricate LSMO/STO/Co heterostructures. After deposition, the Co counter-electrode was etched in a oxygen plasma in order to form a thin layer of antiferromagnetic CoO. The structure was finally capped by a 30 nm Au layer.

The LSMO/LAO/LSMO structure was observed in cross-section transmission electron microscopy (TEM) and a representative image is shown in figure 1a. The surface of the top LSMO layer is very flat and the interface between the bottom LSMO layer and the STO substrate is sharp. However the contrast between LAO and LSMO is weak so that the LAO layer cannot be clearly distinguished. On the high-resolution image of figure 1b, the top and bottom interfaces between the LSMO layers and the LAO spacer are visible, and appear as sharp discontinuities. The LAO thickness corresponds exactly to 7 unit-cells, i.e. ~2.8 nm. No structural defects such as dislocations could be detected in any of the layers and the whole structure can be considered as virtually single-crystalline. We do not discuss the structural properties of the LSMO/LAO/Co sample here but just mention that from X-ray diffraction measurements (not shown), the LSMO and LAO layers are found epitaxial, and the Co layer is hexagonal close packed (hcp) with a (0001) texture.

The LSMO/LAO/LSMO and the LSMO/LAO/Co samples were patterned into micron-sized tunnel junctions using the optical lithography process described in reference [13]. The transport properties were measured in a four-probe configuration in a cryostat equipped with a 6 kOe electromagnet. For all measurements, the electrode resistance was small enough to ensure homogeneous current flow through the junction. Positive bias voltage was chosen to reflect tunneling of electrons from top to bottom electrode.

In figure 2a we plot a typical R(H) curve measured at 4K and a bias voltage $V_{DC}$=10 mV on a 12 µm² tunnel junction. The TMR reaches 300 %, which corresponds to a spin-polarization of 77% for the LSMO/LAO interface. This is somehow lower than the best values obtained for the LSMO/STO interface [13], but larger than those reported for LSMO/TiO$_2$ [14]. The resistance-area (RA) product for these junctions is ~800 kΩ.µm² i.e. larger by a factor of 5-10 than that of similar junctions with STO barriers having the same thickness. This therefore indicates a larger barrier height for LAO than for STO, as expected from the larger band gap of LAO (5.1 eV [7] vs 3.2 eV for STO [15]).

With a spin-polarization of 77% at low temperature, the LSMO/LAO interface can be used as a good spin-dependent analyser of the current tunneling from a Co electrode in LSMO/LAO/Co junctions. A typical R(H) curve measured on this kind of structure is shown on figure 2b, at 4K, $V_{DC}$=10mV and after field-cooling. A clear negative TMR of -11% is obtained. The cycle is asymmetric indicating efficient exchange-biasing of the CoO layer onto the Co electrode. Remarkably, the switching fields between the P and AP configuration in the R(H) curve match very well the reversal fields detected in the M(H) hysteresis cycle, measured prior to patterning (see figure 2c). As the spin-polarization of the LSMO/LAO interface is positive, this negative TMR indicates a negative spin-polarization for the Co/LAO interface, of about $P_{Co/LAO} \approx$ -7% at $V_{DC}$=10 mV. The bias dependence of the TMR for two representative LSMO/LAO/Co junctions is shown on figure 3. The TMR is negative close to zero-bias, and its absolute value increases when increasing bias voltage towards negative values, showing a maximum at about $V_{DC}$ = -0.30 eV. Considering the spin-polarization of the LSMO/LAO interface at this bias (deduced from R(H) curves on LSMO/LAO/LSMO junctions) we find $P_{Co/LAO} \approx$ -18% at -0.30V. Beyond this maximum, |TMR| decreases smoothly, to vanish at $V_{DC} \approx$ 2V. In the positive bias range, the absolute value of the TMR decreases from zero bias up to about 0.25 eV. Then the TMR changes sign and becomes positive, showing a maximum close to 0.5 eV. The TMR finally decreases to cancel at about $V_{DC}$ = 2V.

Remarkably, this peculiar TMR($V_{DC}$) dependence observed in LSMO/LAO/Co junctions is qualitatively very similar to that previously measured in LSMO/STO/Co junctions [16], which brings some insight on the role of the barrier and barrier-electrode interface bonding on tunneling. The negative TMR on this latter system, reflecting some negative spin-polarization for Co at the interface with STO, was ascribed to the tunneling of d-states through STO, in contrast to the tunneling of s-states through $Al_2O_3$. Note that LAO is a perovskite, like STO, but contains Al ions, like $Al_2O_3$ with which the spin-polarization of Co is positive [17,18].

To explain the negative spin-polarization of the STO-Co interface, Oleynik et al performed theoretical calculations and predicted the presence of a small magnetic moment on the Ti atoms (antiferromagnetically coupled to that of the Co atoms) in the last $TiO_2$ sub-layer at the STO-fcc(001) Co interface [19]. However, Co appears to grow in hcp structure and (0001)-oriented on perovskites and this magnetic moment on Ti was never detected experimentally [20,21]. Furthermore, in LAO the ion occupying the B site is Al that has no empty d shells that could hybridize with the orbitals of Co. All these considerations lead us to consider alternative scenarios to explain the striking similarities of the LSMO/STO/Co and LSMO/LAO/Co systems and the negative spin-polarization of both the STO-Co and LAO-Co interfaces.

In MTJs, the tunneling current is known to depend predominantly on two factors, i.e. the electronic structure of the barrier (as illustrated by recent results on MgO-based MTJs [22,23]) and the electronic properties of the interfaces (bonding effects at electrode-barrier interfaces [24], presence of resonant states, etc). To assess the role of the former factor, one has to know the complex electronic band structure of the barrier material. It has been calculated for STO [25] but no such data are available for LAO. Nevertheless, we can compare the real band structure of these two materials. In STO, the conduction bands at the $\Gamma$ point have $\Delta_5$ and $\Delta_{2'}$ symmetry [26], corresponding to unoccupied Ti $t_{2g}$ states. Higher in energy lie Ti $e_g$ states with $\Delta_1$ and $\Delta_2$ symmetry. In their calculations of the complex band structure of STO, Bellini et al find that $\Delta_1$ and $\Delta_5$ states are decaying slowly while $\Delta_2$ states decay more rapidly and $\Delta_{2'}$ states decay the fastest [25]. As previously mentioned, $\Delta_1$ states should correspond to a large barrier height and $\Delta_5$ states to a low barrier height but states of these two symmetries should carry the most important part of the tunneling current. In LAO, the conduction bands at the $\Gamma$ point correspond to lower unoccupied La 4d states [12] split by the dodecahedral crystal field at the perovskite A site, i.e. $d_{z^2}$ and $d_{x^2-y^2}$ states, with $\Delta_1$ and $\Delta_2$ symmetry, respectively. If we assume the same hierarchy of decay rates in LAO as in STO, $\Delta_1$ states should contribute predominantly to tunneling as they would have a small decay rate and in that case also correspond to a small barrier height.

The matching of symmetry of the wave functions between the electrodes and the barrier also has to be considered. In LSMO, the electronic structure close to $E_F$ is relatively simple, with only spin-up $\Delta_1$ and $\Delta_2$ metallic bands [27]. From this analysis, we thus expect the tunneling current in LSMO/STO/Co and LSMO/LAO/Co to be predominantly carried by $\Delta_1$ states, at least at low bias. Now, what is required is the nature of the states available at $E_F$ in the Co electrode. Even if the electronic structure of hcp Co along the (0001) direction (ΓA) has been calculated [28], it cannot be used directly since bonding effects between Co and O at the LAO/Co interface will influence the symmetry of the electronic states and must therefore be taken into account. Indeed, recent calculations of fcc(111)-Co/(0001)-$Al_2O_3$ interfaces have demonstrated that such bonding effects deeply alter the electronic structure and can even reverse the sign of the spin-polarization [18]. Following this observation, a way to understand the similarity between the LSMO/LAO/Co and LSMO/STO/Co bias dependences is to consider that the hybridization effects occurring between Co atoms in a hcp(0001) plane and the O atoms in both perovkistes (located at virtually identical positions in LAO and STO) are similar. A detail structural analysis of the LAO/Co interface is in progress but in LSMO/STO/Co junctions, the STO layer was found to be $TiO_2$ terminated, with some evidence for CoO-type bonding [29]. To push the analysis further, calculations of the band structure of $BO_2$/hcp(0001)-Co interfaces are thus required. Together with the complex band structure of LAO, these calculations may allow to quantitatively understand the full TMR($V_{DC}$) dependence.

In summary, we have successfully used epitaxial layers of the promising high-k dielectric $LaAlO_3$ as tunnel barriers in magnetic tunnel junctions. Spin-dependent transport measurements on LSMO/LAO/LSMO junctions indicate a low-temperature spin-polarization of 77% for the LSMO/LAO interface. On the contrary, the spin-polarization of Co at the interface with LAO is negative, as occurs with STO. Our TMR results combined with the large band offset between LAO and Si [5], makes LAO a suitable barrier material for spin-injection into Si. We propose that the striking similarities between the LSMO/LAO/Co and LSMO/STO/Co systems are mostly related to specific bonding effects occurring between the hcp(0001) Co planes and the O ions in the perovskite barrier material. Input from theory on this point and on the complex band structure of these perovskite oxides is required for a better understanding.

* corresponding author : manuel.bibes@ief.u-psud.fr

Figures

Fig 1: (a) Transmission electron microscopy cross-section in [010] zone of a LSMO/LAO/LSMO trilayer. (b) Enlargement of region squared in (a). In the imaging conditions used, MnO and AlO atomic columns are bright. The directions indicated refer to the pseudocubic perovskite unit cell.

Fig 2 : Field dependence of the resistance, at 10 mV and 4K for a LSMO/LAO/LAMO junction (a) and a LSMO/LAO/Co junction, after field cooling (b). (c) Magnetization hysteresis cycle at 10K for a LSMO/LAO/Co trilayer, after field cooling. The field was applied in-plane.

Fig 3 : Bias-dependence of the TMR for two LAMO/LAO/Co junctions. Positive bias corresponds to electrons tunneling from Co to LSMO.

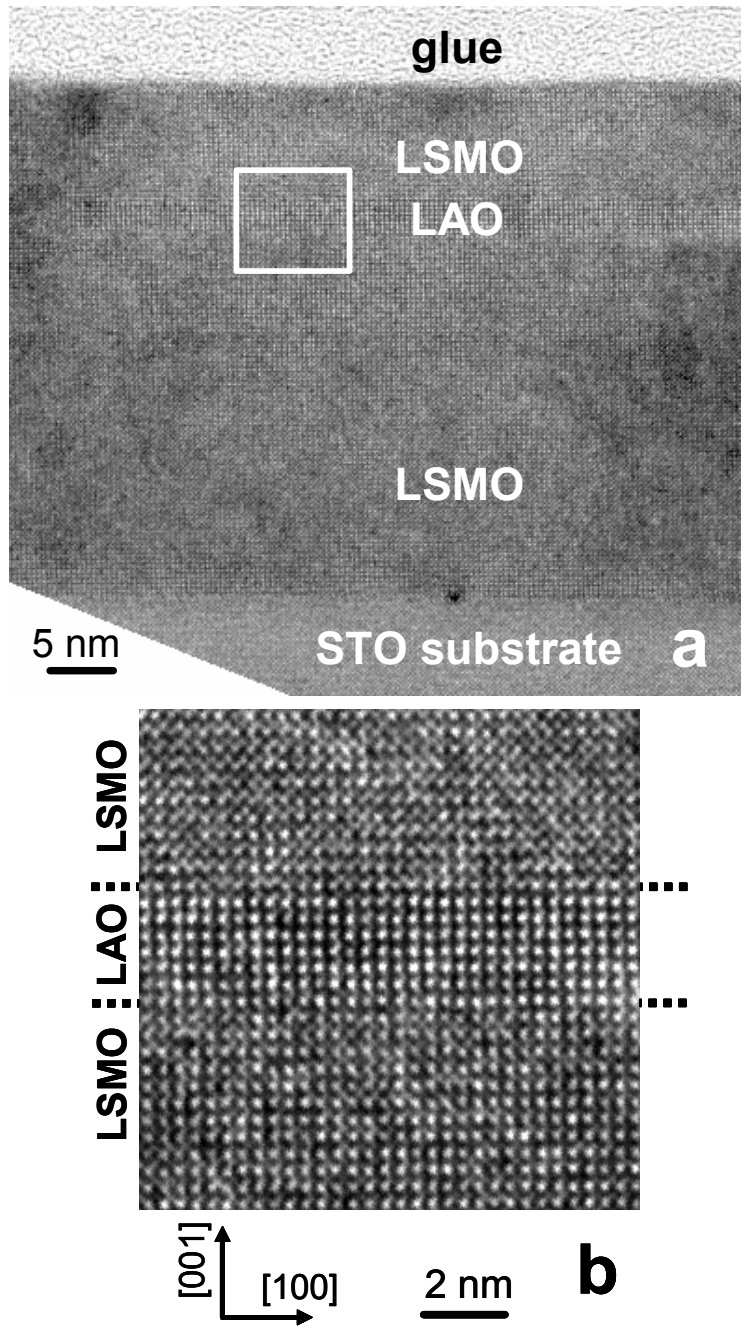

Fig 1

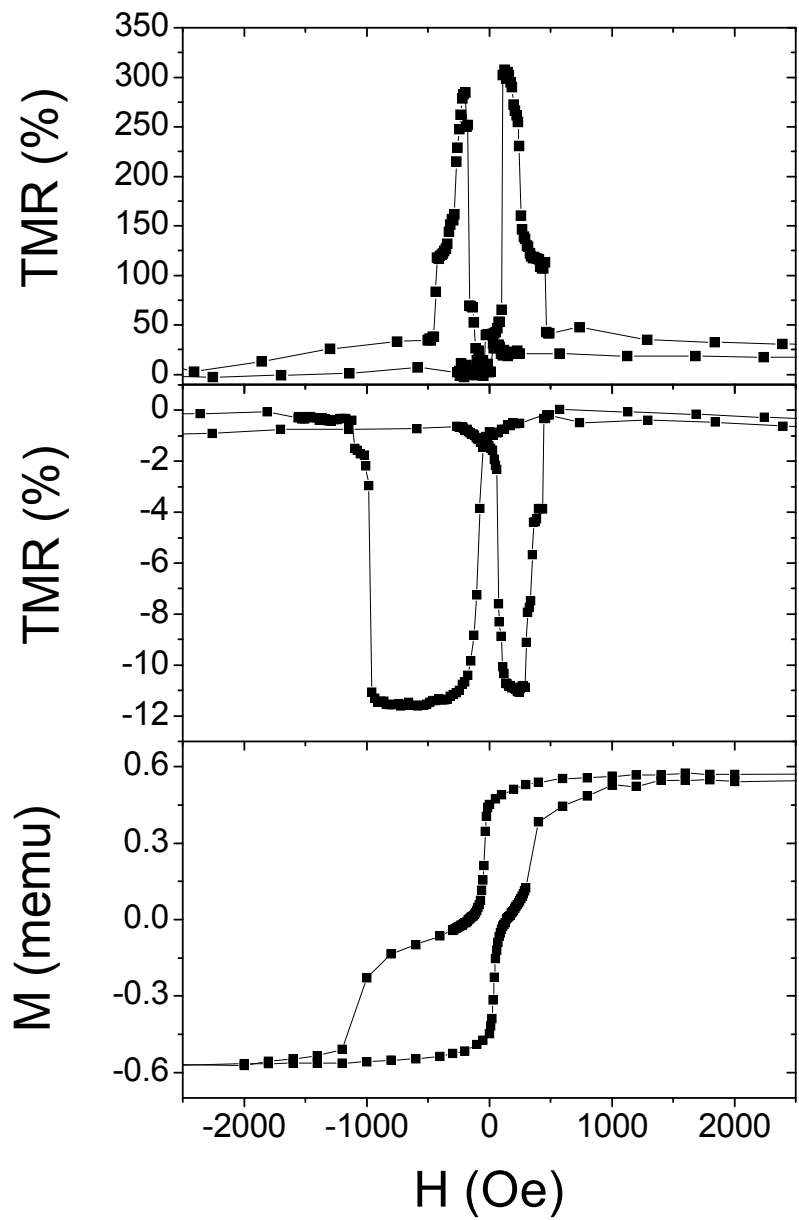

Fig 2

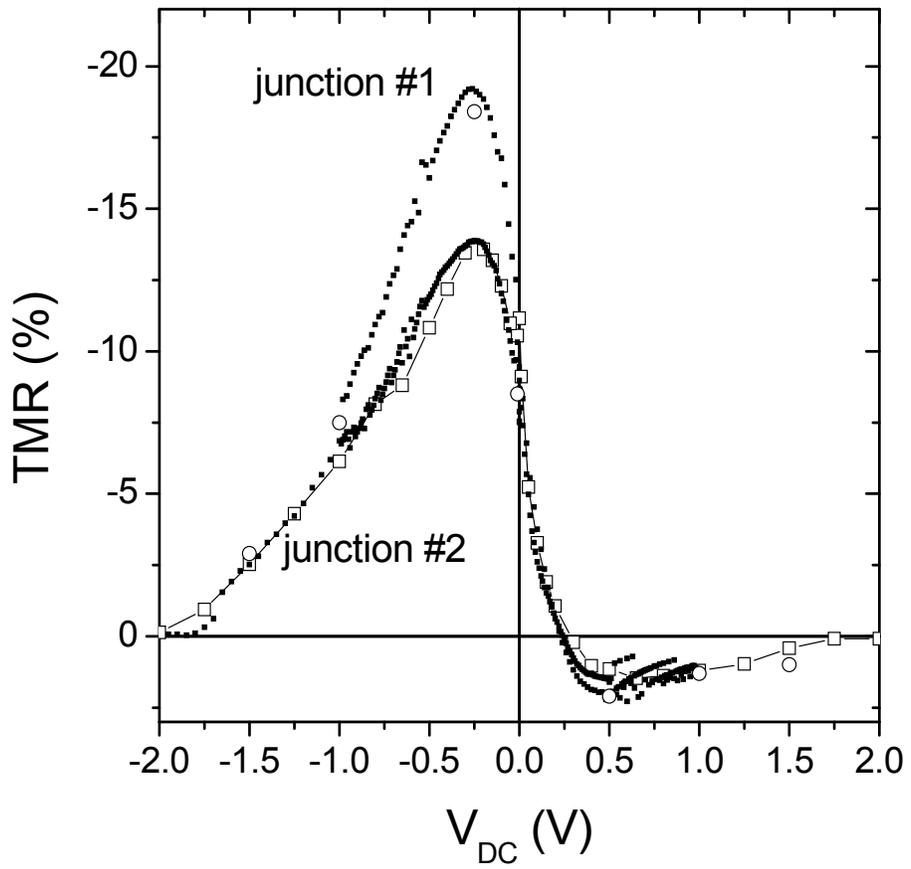

Fig 3